\documentclass[usenatbib]{mnras}
\usepackage[T1]{fontenc}
\usepackage{aecompl}
\usepackage{amsmath,amsfonts,amssymb}
\usepackage[pdftex]{graphicx}
\usepackage{aas_macros}
\usepackage[usenames]{xcolor}
\usepackage{color}
\usepackage{textcomp,gensymb}
\usepackage{times}
\usepackage{subfig}
\usepackage{ulem}

\newcommand{\kms}{${\rm km\,s}^{-1}$}

\newcommand{\rr}[1]{\textcolor{black}{#1}}

\title[Observing flybys in protoplanetary discs]{Flybys in protoplanetary discs --- II. Observational signatures}

\author[N. Cuello et al.]{\parbox{\textwidth}{
Nicol\'as Cuello,$^{1,2,6}$\thanks{corresponding author: cuellonicolas@gmail.com}
Fabien Louvet,$^{3}$
Daniel Mentiplay,$^{4}$
Christophe Pinte,$^{4,5}$\\
Daniel~J. Price,$^{4}$
Andrew J. Winter,$^{6}$
Rebecca Nealon,$^{6}$
Fran\c cois M\'enard,$^{5}$\\
Giuseppe Lodato,$^{7}$
Giovanni Dipierro,$^{6}$
Valentin Christiaens,$^{4}$
Mat\'ias Montesinos,$^{8,2,9}$
Jorge Cuadra,$^{1,2}$
Guillaume Laibe,$^{10}$
Lucas Cieza,$^{11}$
Ruobing Dong,$^{12}$\\
Richard Alexander$^{6}$}
\vspace{0.2cm}\\
$^{1}$Instituto de Astrof\'isica, Pontificia Universidad Cat\'olica de Chile, Santiago, Chile,\\
$^{2}$N\'ucleo Milenio de Formaci\'on Planetaria (NPF), Chile,\\
$^{3}$Departamento de Astronom\'ia de Chile, Universidad de Chile, Santiago, Chile,\\
$^{4}$Monash Centre for Astrophysics (MoCA) and School of Physics and Astronomy, Monash University, Clayton VIC 3800, Australia,\\
$^5$Univ. Grenoble Alpes, CNRS, IPAG, F-38000 Grenoble, France,\\
$^{6}$School of Physics and Astronomy, University of Leicester, University Road, Leicester, LE1 7RH, United Kingdom,\\
$^{7}$Dipartimento di Fisica, Universit\`a Degli Studi di Milano, Via Celoria, 16, Milano, 20133, Italy,\\
$^{8}$Universidad de Valparaiso, Chile,\\
$^{9}$Chinese Academy of Sciences South America Center for Astronomy, National Astronomical Observatories, CAS,
Beijing 100012, China,\\
$^{10}$Univ Lyon, Univ Lyon1, Ens de Lyon, CNRS, Centre de Recherche Astrophysique de Lyon UMR5574, F-69230, Saint-Genis-Laval, France,\\
$^{11}$Facultad de Ingenier\'ia y Ciencias, N\'ucleo de Astronom\'ia, Universidad Diego Portales, Av. Ejercito 441. Santiago, Chile,\\
$^{12}$Department of Physics \& Astronomy, University of Victoria, Victoria, BC, V8P 1A1, Canada.}

\begin{document}
\date{Accepted 2019 October 15. Received 2019 October 11; in original form 2019 August 23}

\pagerange{\pageref{firstpage}--\pageref{lastpage}} \pubyear{2019}

\maketitle

\label{firstpage}

\begin{abstract}
Tidal encounters in star clusters perturb discs around young protostars. In \citet[][Paper I]{Cuello+2019b} we detailed the dynamical signatures of a stellar flyby in both gas and dust. Flybys produce warped discs, spirals with evolving pitch angles, increasing accretion rates, and disc truncation. Here we present the corresponding observational signatures of these features in \rr{optical/near-infrared} scattered light and \rr{(sub-) millimeter} continuum and CO line emission. Using representative prograde and retrograde encounters for direct comparison, we post-process hydrodynamical simulations with radiative transfer methods to generate a catalogue of multi-wavelength observations. This provides a reference to identify flybys in recent near-infrared and sub-millimetre observations \rr{(e.g., RW~Aur, AS~205, HV~Tau \& DO~Tau, FU~Ori, V2775~Ori, and Z~CMa)}.
\end{abstract}



\section{Introduction}
\label{sec:intro}

Protoplanetary discs are the cradle of newborn planets. Given typical disc lifetimes --- 1 to 10 Myr --- planets should form within these systems in less than a few Myr. However, despite an active search for embedded planets in protoplanetary discs, only three candidates have been reported so far: in PDS~70 \citep{Keppler+2018,Muller+2018}, HD~163296 \citep{Pinte+2018} and HD~97048 \citep{Pinte+2019}. A complete theoretical understanding of planet formation remains elusive \citep{Armitage2018}. 

Numerous radial and azimuthal features such as spirals \citep{Benisty+2015, Benisty+2017,Perez+2016,Huang+2018}, shadows \citep{Avenhaus+2014, Stolker+2016, Benisty+2018}, gaps \citep{ALMA+2015, Tsukagoshi+2016, Dipierro+2018,Andrews+2018}, warps \citep{Langlois+2018,Casassus2018a,vanderPlas+2019}, horseshoes \citep{vanderMarel+2013,Boehler+2017}, and clumps \citep{Dong+2018,Gratton+2019} have been reported. Such structures are potential signposts of disc--companion interactions \citep[for instance]{Dong+2015a,Price+2018,Poblete+2019,Calcino+2019}.

Spirals and misaligned inner/outer discs are often assumed to be indicators of massive (planetary or stellar) perturbers. These companions can either be external \citep{Clarke&Pringle1993,Pfalzner2003,Quillen+2005,Dong+2015a} or internal \citep{Facchini+13,LodatoFacchini13,Cazzoletti+17,Aly+18,Price+2018,Keppler+2018,Cuello&Giuppone2019} to the protoplanetary disc. Interestingly, planetary companions on inclined orbits are able to warp the disc \citep{Facchini+14,Nealon+2018} and produce observable features in scattered light \citep{Zhu2019,Nealon+2019}. Massive companions may also lead to disc breaking with observational signatures present in both scattered light and mm wavelengths \citep{Facchini+2018,Montesinos&Cuello2018,Cuello+2019a}. Additionally, accretion from an external envelope \citep{Harsono+2011,Lesur+2015,Hennebelle+2017} and chaotic interactions within a molecular cloud \citep{Bate+10,Bate2018} can also form spirals and misaligned discs. In this work, we focus on the scenario where a disc is perturbed by a stellar companion on an unbound (parabolic) orbit. Our aim is to predict the resulting observational signatures.

Parabolic star-disc encounters are expected to occur during the early phases of stellar evolution (< 1~Myr) in clustered associations of stars \citep{Craig&Krumholz2013, Pfalzner2013, Winter+2018b}. It is therefore likely that at least one of the stars involved in the encounter has a protoplanetary disc. Signatures of such encounters have been observed, e.g. in RW~Aur \citep{Dai+2015,Rodriguez+2018}, HV \& DO~Tau \citep{Winter+2018c}, FU~Ori \citep{Takami+2018}, and AS~205 \citep{Kurtovic+2018}. Provided the encounter is close enough, the stellar flyby can dramatically affect the disc structure creating spirals, bridges, warps, and diffuse nebulae. In a previous study \citep[hereafter Paper~I]{Cuello+2019b}, we examined the dynamical signatures of flybys in the gas and in the dust.

Our aim in this paper is to provide observational diagnostics of protoplanetary discs experiencing a stellar flyby. We investigate the disc emission at different wavelengths as a function of orbital inclination of the flyby. In Section~\ref{sec:flybysynthobs}, we describe the methodology followed to perform the radiative transfer calculation and the corresponding synthetic observations. In Section~\ref{sec:synthobs}, we classify the flyby signatures at different wavelengths in order to provide a guide to interpret recent observations. Our results are discussed in the light of recent observations of interacting stellar objects in Section~\ref{sec:discussion}. We conclude in Section~\ref{sec:conclusions}.


\section{Methods}
\label{sec:flybysynthobs}

In Paper~I, we presented a series of smoothed particle hydrodynamical (SPH) simulations of a protoplanetary disc disturbed by a single stellar flyby. We used {\sc Phantom} \citep{Price+2017}. Calculations were performed for a range of orbital (prograde and retrograde) inclinations and grain sizes (Sect.~\ref{sec:model}). The response of the protoplanetary disc to the gravitational perturbation of the intruder depends on the orbital inclination and stellar mass ratio. For instance, disc truncation is more efficient for prograde encounters; whereas disc warping is greater in retrograde encounters \citep{XG2016}. Both effects increase with increasing pertuber-to-host mass ratio, $q$. Here, we characterise the observational signatures of those encounters by post-processing the hydrodynamical simulations with {\sc mcfost} \citep{Pinte+2006,Pinte+2009}, see radiative transfer calculations in Sect.~\ref{sec:mcfost}.


\subsection{Disc model and flyby parameters}
\label{sec:model}

We considered a $1\,M_\odot$ star surrounded by a protoplanetary disc with mid-plane initially in the $xy$- (or equivalently $z=0$) plane. We set up a disc with an initial inner and outer disc radius of $R_{\rm in}=10$ au and $R_{\rm out}=150$~au, respectively. At the beginning of the calculation the disc surface density followed a power-law profile $\Sigma \propto R^{-1}$, where $R$ is the cylindrical radius. We adopted a relatively large and massive disc as it corresponds to the kind of systems that can be observed at high spatial resolution with current instrumentation. We modelled the disc using $10^6$ gas SPH particles assuming a total gas mass of $0.01~M_\odot$. We set the SPH viscosity parameter $\alpha_{\rm AV} \approx 0.26$, approximately equivalent to a mean \cite{Shakura&Sunyaev1973} disc viscosity $\alpha_{\rm SS}$ of $0.005$ \citep{Lodato&Price2010}. We used a locally isothermal equation of state where the temperature is a function of radial distance from the disc-hosting star according to $T(r) = 64 \,{\rm K} \,(r/r_{\rm in})^{-1/2}$. This corresponds to a disc scale-height $H/R = 0.05$ at $R = R_{\rm in}$ and $H/R = 0.1$ at $R = R_{\rm out}$, consistent with recent disc observations \citep[e.g.][]{Pinte+2016}. Using an adiabatic equation of state might change the disc response as shown by \cite{Lodato+2007}, but we do not consider such effects in this work.

The dust content of the disc was modelled using two different methods according to the grain size considered: the one-fluid method for micron-sized grains (1-10 $\mu$m) \citep{Price&Laibe2015,Ballabio+2018}, and with dust modelled as a separate set of particles for grains between $100$~$\mu$m and 10~cm \citep{Laibe&Price2012}. We performed calculations for each grain size individually, and stacked these calculations together for radiative transfer post-processing. The dust was initialised to follow the same radial and vertical density profile as the gas, with mass scaled down by a factor of 100 from the gas mass. For further details and tests, see Paper~I (Section~2.2 and Appendices~B and C).

We considered equal-mass encounters ($q=1$). We set the $1\,M_\odot$ perturber on a parabolic orbit with initial separation 10 times the pericentre distance with $R_{\rm peri}=200$ au. Since $R_{\rm peri} > R_{\rm out}$ this implies a non-penetrating flyby. The perturber does not have a disc previous to the encounter. We define $\beta$ to be the angle between the angular momentum vector of the disc and that of the flyby orbit (see figure~1 in Paper~I). When $\beta \neq 0$ there is an additional angle between the direction of pericentre and the line of intersection of the disc and the orbital plane. We ignored this additional angle since $\beta$ dominates the variation in angular momentum transfer for the dominant $m=2$ inner Lindblad resonance in close, non-penetrating encounters \citep{Ostriker1994,Winter+2018a}. Being interested in the 3D disc structure during the flyby, we chose two representative orbits misaligned with respect to the disc mid-plane: inclined prograde ($\beta=45\degr$) and inclined retrograde ($\beta=135\degr$).

In order to build a representative disc made of a mixture of gas and dust of multiple grain sizes, we stacked the distributions of different grain size (0.1~mm, 1~mm, 1~cm and 10~cm) following the procedure outlined in \citet{Mentiplay+2019}. \rr{Since grains with sizes ranging between 1 $\mu$m and 10 $\mu$m are strongly coupled to the gas, we assumed that these follow the gas distribution.} Therefore we ignored those calculations when stacking. We chose the \rr{gas distribution} in the 0.1~mm (gas+dust) calculation as \rr{the reference for the other grain sizes. In other words,} we discarded the gas from the other grain size calculations and added their dust particles to the 0.1~mm calculation. The gas distributions in each individual grain size calculations were found to be similar because of the low dust-to-gas ratio. We detail our procedure for producing synthetic observations from these simulations below.


\subsection{Radiative transfer calculations}
\label{sec:mcfost}

We used the stacked disc models described in Sect.~\ref{sec:model} as input to the radiative transfer code {\sc mcfost} \citep{Pinte+2006,Pinte+2009}. The radiative transfer was calculated on an unstructured Voronoi mesh derived from the SPH gas particles. \cite{Nealon+2019} provide further details on the mesh construction. This allowed us to perform radiative transfer calculations on the complex geometry of the perturbed disc during the flyby, without requiring interpolation between the SPH and radiative transfer codes.

We considered two sources of radiation: the central star surrounded by the disc, and the perturber. This combination asymmetrically illuminates the disc (see Figs.~\ref{fig:scatteredlight-pro} and \ref{fig:scatteredlight-retro}). Considering each star has a mass of $1~\mathrm{M}_{\odot}$, we used a stellar spectrum and luminosity derived from a 3~Myr Siess isochrone \citep{Siess+2000}: $\mathrm{T_{eff}} = 4262~K$, $\mathrm{L} = 0.997~\mathrm{L}_{\odot}$, and $\mathrm{R} = 1.722~\mathrm{R}_{\odot}$. We used $10^7$ photon packets in the temperature calculation and to compute the monochromatic specific intensity.  Final images were generated using a ray-tracing method \citep{Pinte+2009}. Dust optical properties were computed using Mie theory, assuming astro-silicate composition \citep{Draine&Lee1984}. For the radiative transfer, we rescaled the dust mass of each dust size bin in order to obtain a total dust-to-gas ratio of $0.01$. Within each cell of the Voronoi mesh the grain size distribution was split into 100 logarithmically spaced bins. We assumed that the spatial distribution of grains smaller than $1~\mu$m followed the gas, i.e. the spatial distribution of modelled grain sizes did not affect the spatial distribution of small grains. The size distribution of these grains was assumed to follow a power-law $\mathrm{d}n(a) \propto a^{-3.5}\,\mathrm{d}a$. Grains between $1~\mu$m and $0.1$~mm follow the same power law such that the mass in $0.1$~mm grains match the models for that grain size. For grains larger than $0.1$~mm the spatial distribution of dust with respect to the gas was determined from the output from the {\sc Phantom} simulations. \rr{The sizes for} grains between the modelled size bins were interpolated using a linear interpolation in log-log space.

We set the distance to $100$~pc and the image size to $1000\times1000$~au (equivalent to $10^{\prime\prime}\times10^{\prime\prime}$). We calculated scattered light images in the H-band at $1.6~\mu$m, thermal emission at $850~\mu$m, and $^{12}$CO J=3--2 molecular emission. For the CO emission we assumed a CO-to-H${}_2$ abundance ratio of $10^{-4}$ \citep{Lacy+1994,France+2014} and produced channel maps at $0.1$~km/s resolution with a turbulent velocity of $0.05$~km/s. We assumed the CO is in local thermodynamic equilibrium (as we only look at low-J CO lines) and a gas temperature equal to the dust temperature computed by {\sc mcfost}.

For direct comparison with recent observations in the sub-millimetre, we produced synthetic {\sc alma} observations of our models with the {\sc casa} package \citep{McMullin+2007}. We computed the $^{12}$CO(3-2) emission (resp. continuum) at the central frequency of $\sim$346~GHz (resp. 353~GHz) in a bandwidth of 23~MHz (resp. 600~MHz) and a spectral resolution of $\sim$0.1~MHz (resp. 600~MHz). All the synthetic observations were done using the ``alma.cycle10.cfg'' configuration of the interferometer and a precipitable water vapour of 0.4\,mm to set the thermal noise. This resulted in a synthetic beam size of $\sim$0.30'' for both the $^{12}$CO and continuum emissions (see Figs.~\ref{fig:continuum} and \ref{fig:CO-channels}).

\section{Results}
\label{sec:synthobs}
We adopt the observational convention where North is up and East is left for describing the synthetic images obtained at different wavelengths.


\subsection{Scattered light}
\label{sec:scattered}

\begin{figure*}%
    \centering
    {\includegraphics[width=0.95\textwidth]{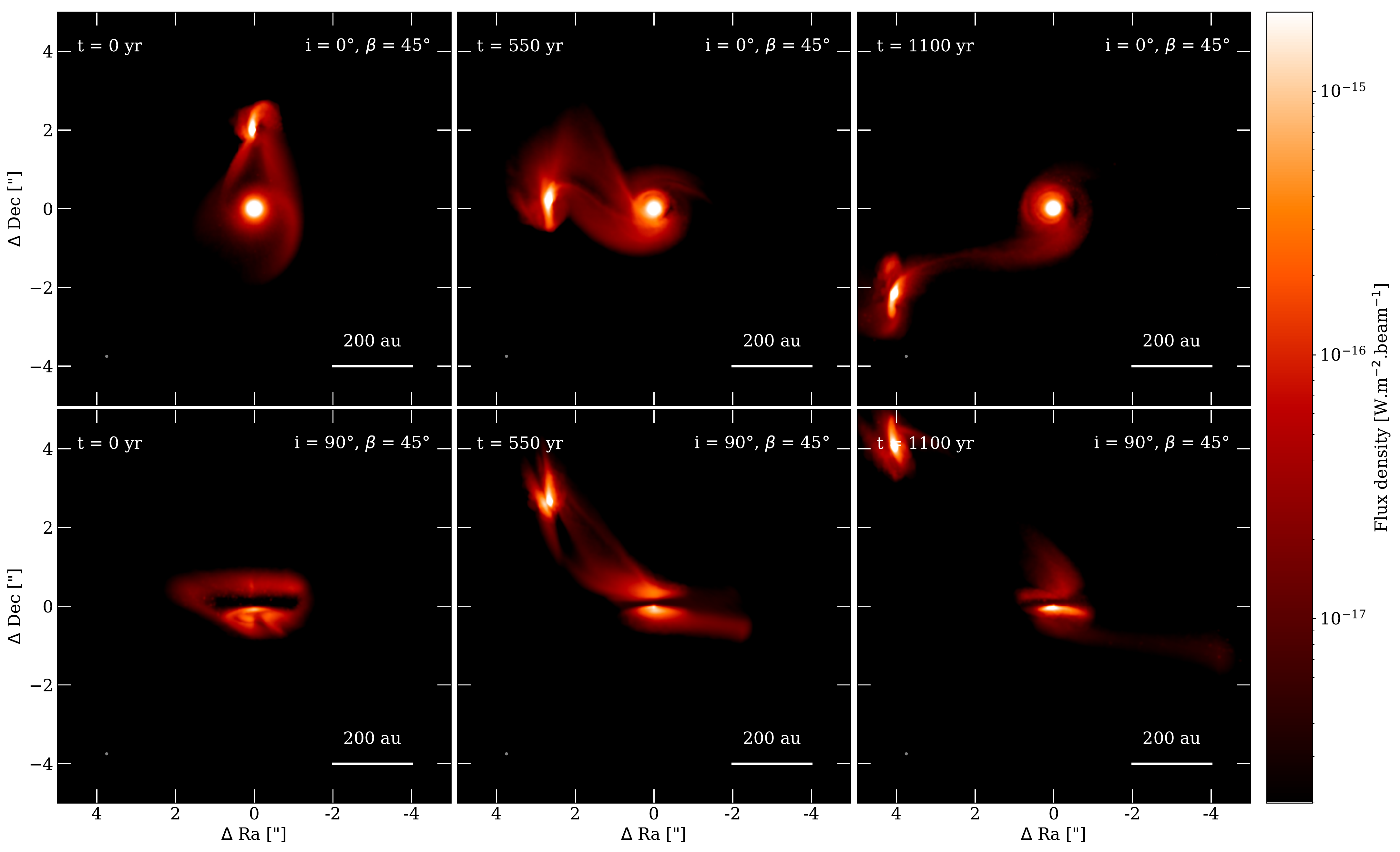}}%
    \caption{Disc evolution during an inclined prograde flyby ($\beta=45\degr$) in scattered light ($\lambda=1.6 \,\mu$m). A prominent bridge of material appears between the perturber and the disc. Upper row shows face-on view ($i=0\degr$) while lower panel shows edge-on ($i=90\degr$). The beam is 50~mas~$\times$~50~mas, and is indicated by the small grey circle in the bottom left of each figure.}%
    \label{fig:scatteredlight-pro}%
\end{figure*}

\begin{figure*}%
    \centering
    {\includegraphics[width=0.95\textwidth]{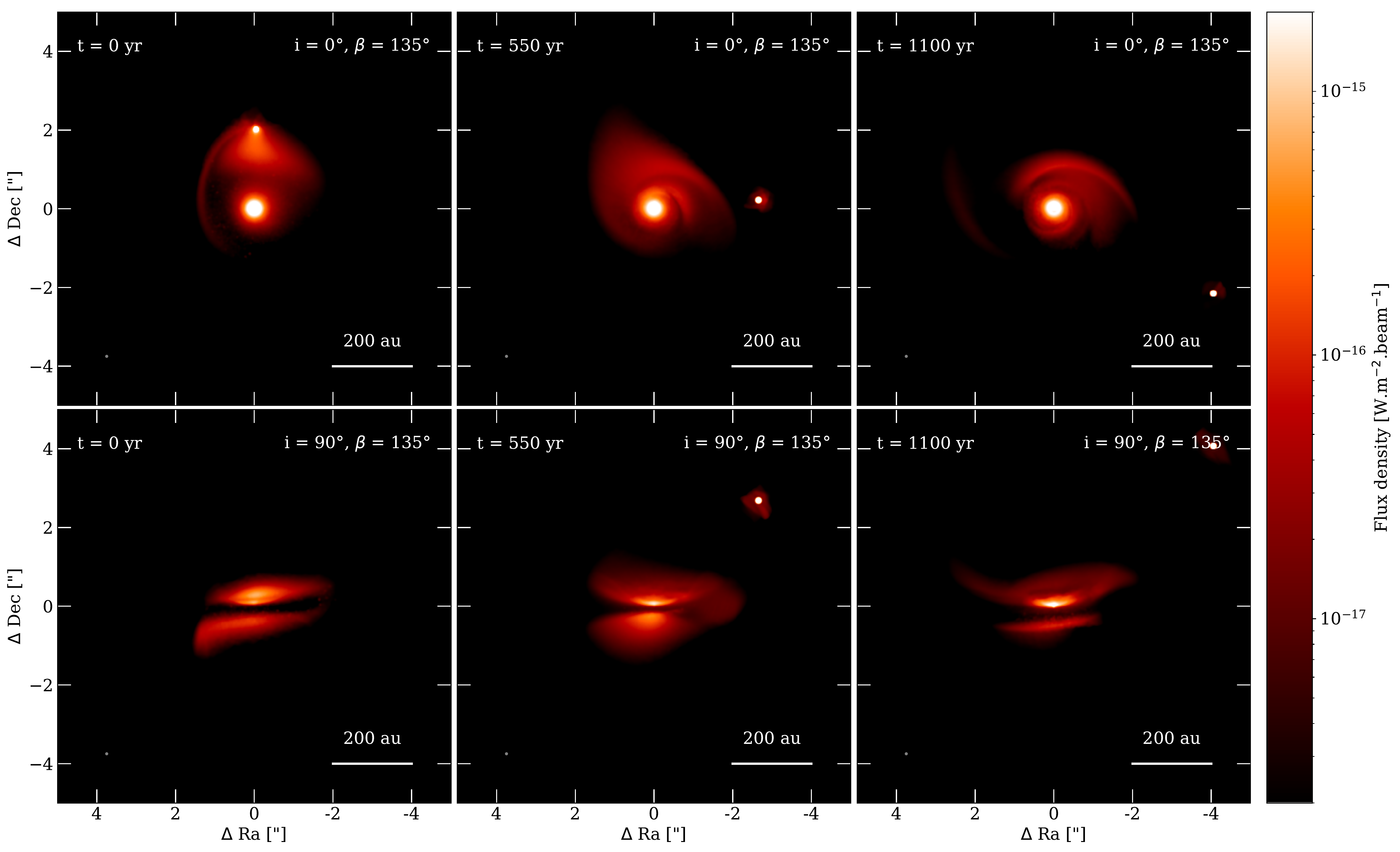}}
    \caption{Same as Fig.~\ref{fig:scatteredlight-pro} but for an inclined retrograde flyby ($\beta = 135\degr$). We observe a spiral arm to the North. The beam is 50~mas~$\times$~50~mas, and is indicated by the small grey circle in the bottom left of each figure.}%
    \label{fig:scatteredlight-retro}%
\end{figure*}

Figures~\ref{fig:scatteredlight-pro} and \ref{fig:scatteredlight-retro} show scattered light synthetic observations at $\lambda~=~1.6~\mu$m for $\beta=45\degr$ and $\beta=135\degr$, respectively. We show three different times: when the perturber is at pericentre ($t=0$), shortly after ($t=550$~yr) and just prior to the perturber leaving the field ($t=1100$~yr). The three-dimensional distance between the stars is 200, 375, and 612 au (respectively). The disc is shown face-on ($i=0\degr$, top rows) and edge-on ($i=90\degr$, bottom rows) from the observer's perspective relative to the initial disc mid-plane.

The main observational difference between the prograde and retrograde encounter is the appearance of prominent spirals in the prograde case, as expected from theory \citep{Ostriker1994,Winter+2018a}. These are observed shortly after pericentre passage (middle frames in Fig.~\ref{fig:scatteredlight-pro}) where one spiral arm appears between the two stars, and the other anchored in the disc on the opposite side. Both spirals lie out of the initial disc plane. 

Interestingly, the spiral arm lying between the two stars is brighter since it is simultaneously illuminated by both stars. This interstellar {\it bridge} remains at later evolutionary stages (see $t=1100$ yr in the right frames of in Fig.~\ref{fig:scatteredlight-pro}). Since such features are not observed for retrograde flybys, gas bridges observed in scattered light imply prograde encounters. In principle similar bridges could also be caused by an outer bound companion, although for an unbound encounter the bridge is expected to \rr{extend over longer distances}. We discuss this further in Sect.~\ref{sec:statflybys}.

Stellar flybys also induce warps. In contrast to bridges, warps are more apparent in discs which have undergone retrograde encounters. This may be observed by comparing the disc edge-on views for $\beta=45\degr$ and $\beta=135\degr$ at $t=1100$~yr (bottom right frames in Figs.~\ref{fig:scatteredlight-pro} and \ref{fig:scatteredlight-retro}). For the retrograde flyby, \rr{the disc --- initially exactly edge-on --- develops} an asymmetric illumination between the upper and the lower parts \rr{with respect to the disc mid-plane. This is due to the disc warping caused by the flyby. As a result, after the encounter, both the far and near sides of the upper half of the disc can be seen}. For instance, at $t=1100$~yr, different illuminations between North and South exist in both the prograde and the retrograde cases. Hence, significant disc warping can be considered as a signpost of a flyby. We discuss the effect of more massive perturbers and more penetrating encounters in Sect.~\ref{sec:suspect}.

In line with previous works \citep{Jilkova+2016, Breslau+2017, Winter+2018a}, we find that \rr{perturbers on prograde orbits} are more efficient at acquiring material from the circumprimary disc than their retrograde counterparts. Also, since prograde encounters unbind material from smaller radii in the disc, the truncation radius is smaller for such encounters. However, the outer radius of a disc prior to encounter is not an observable quantity. Therefore, the extent of the disc in scattered light alone does not constrain encounter orientation.

All of our scattered light synthetic observations in Figures~\ref{fig:scatteredlight-pro}~and~\ref{fig:scatteredlight-retro} closely resemble the corresponding gas density fields shown in figures~2 and 3 of Paper~I. This occurs because micron-sized grains are well coupled to the gas. On the other hand, the unique shadowing patterns due to light obstruction by the disc material are only captured through radiative transfer calculations. These effects may help to constrain the disc inclination when dust continuum and kinematics are not available.

\begin{figure*}%
    \centering
   \includegraphics[width=0.75\textwidth]{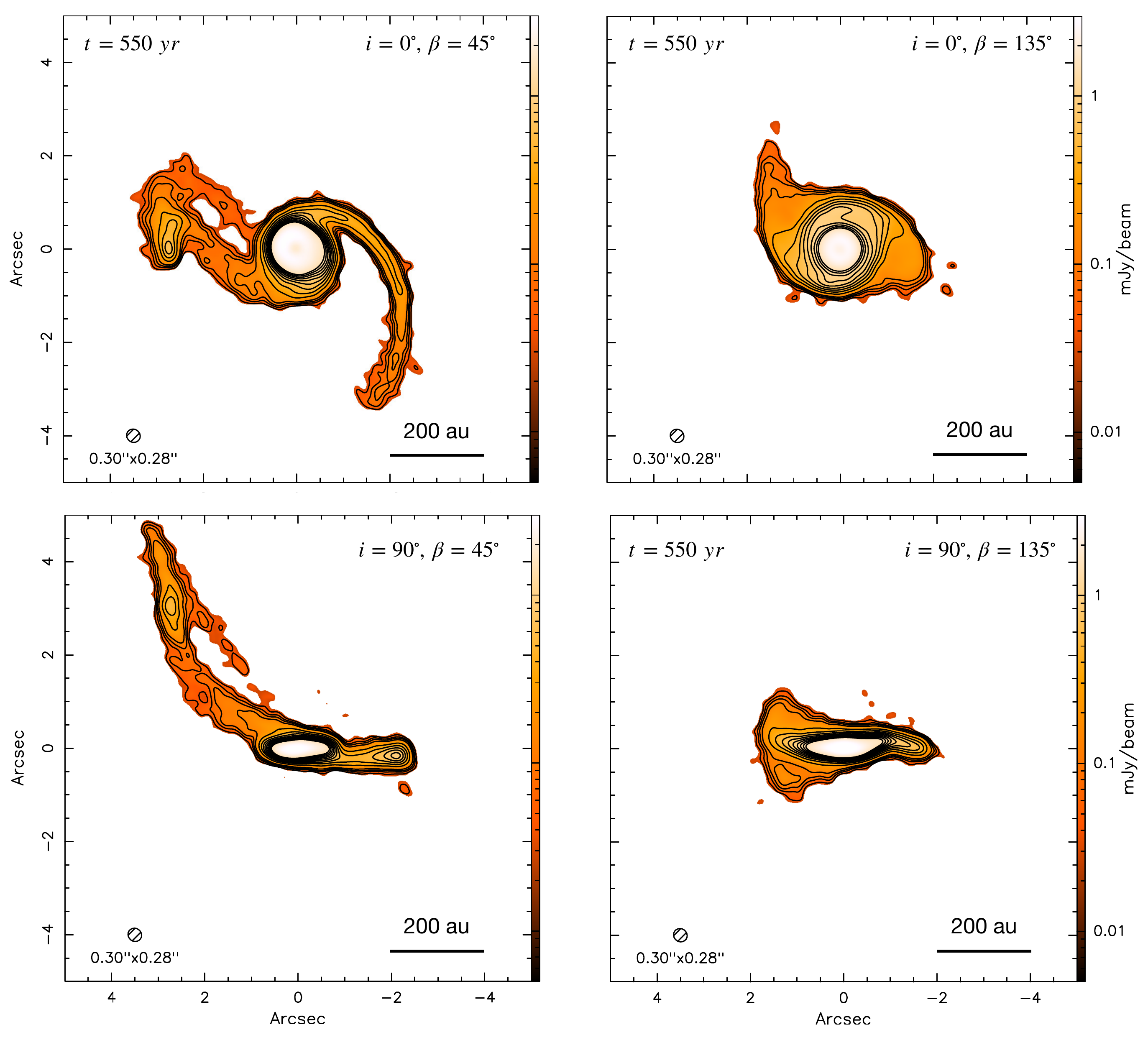}%
    \caption{ALMA Band 7 synthetic observations of the thermal dust emission of the system $550$~yr after the passage at pericentre for $\beta=45\degr$ (left column) and $\beta=135\degr$ (right column). The synthetic observations mimic 2 hours of observation and are to be compared to the scattered light images in the middle columns of Figs.~\ref{fig:scatteredlight-pro} and \ref{fig:scatteredlight-retro}. The top row shows the synthetic observation of the system inclined by $i=0\degr$ (face-on view) while the bottom row shows the system inclined by $i=90\degr$ (edge-on view). The contours show the level at 3, 5, 7 $\sigma$ and then 10 $\sigma$ to 100 $\sigma$ by step of 10 $\sigma$, with $\sigma=0.014$~mJy/beam. To ease the visualisation, the data are clipped at 2$\sigma$.}%
    \label{fig:continuum}%
\end{figure*}


\subsection{Continuum \rr{thermal} emission}
\label{sec:continuum}

Figure~\ref{fig:continuum} shows the disc emission at $850~\mu$m for $\beta=45~\degr$ (\rr{left} column) and for $\beta=135~\degr$ (\rr{right} column) at $t=550$~yr. The spiral in the East in the left panels of Fig.~\ref{fig:continuum} corresponds to the interstellar bridge seen in scattered light (see middle column in Fig.~\ref{fig:scatteredlight-pro}). Here we show observations in face-on and edge-on configurations. Intermediate viewing inclinations (not shown) cover a broader range of pitch angles and morphologies. Shortly after periastron passage, spirals exhibit large pitch angles ($\sim30\degr$ or more) which decrease with time. Because of projection and dynamical effects \citep{Pfalzner2003}, the two spirals do not necessarily have the same pitch angle.

For the $\beta=135\degr$ case, two symmetrical spirals appear shortly after the passage at pericentre ($t=550$~yr). These spirals appear less prominent and more compact with respect to those induced by the $\beta=45\degr$ encounter. As the distance between the stars increases with time, the spirals quickly disappear as the gas in the disc recircularises. This happens in a few orbital periods at the spiral location. The continuum emission seen edge-on shows a more warped geometry than the prograde encounter. This is because inclined retrograde encounters are more efficient at tilting and twisting the disc \citep{XG2016,Cuello+2019b}.

For the non-penetrating encounters considered ($R_{\rm peri}/R_{\rm out}\approx1.3$), no disc material is captured by the perturber for retrograde flybys; whereas a circumsecondary disc forms for prograde flybys. This is seen in the scattered light and continuum images. However, for more evolved discs, we expect more compact dust distributions for mm-sized grains because of radial drift (see figures~6 and 7 in Paper~I). This translates into more compact spirals arms in the continuum for systems where the flyby occurs after several Myr. Therefore, for more evolved discs, the perturber only captures gas, leaving the millimetre\rr{-sized} dust unaffected. This scenario is further discussed in Sect.~\ref{sec:flybyobs} based on recent observations of interacting stellar objects. Our main result is that flyby-induced spirals (if present) are in principle detectable with {\sc alma} in Band~7 for a reasonable integration time ($\sim$2~h). \rr{Other bands (e.g. Band~3) and more extended configurations (i.e. higher resolution $\sim 0.1$ arcsec) would require longer integration times.}


\subsection{CO kinematics}
\label{sec:CO}

\begin{figure*}%
    \centering
   \includegraphics[width=0.95\textwidth]{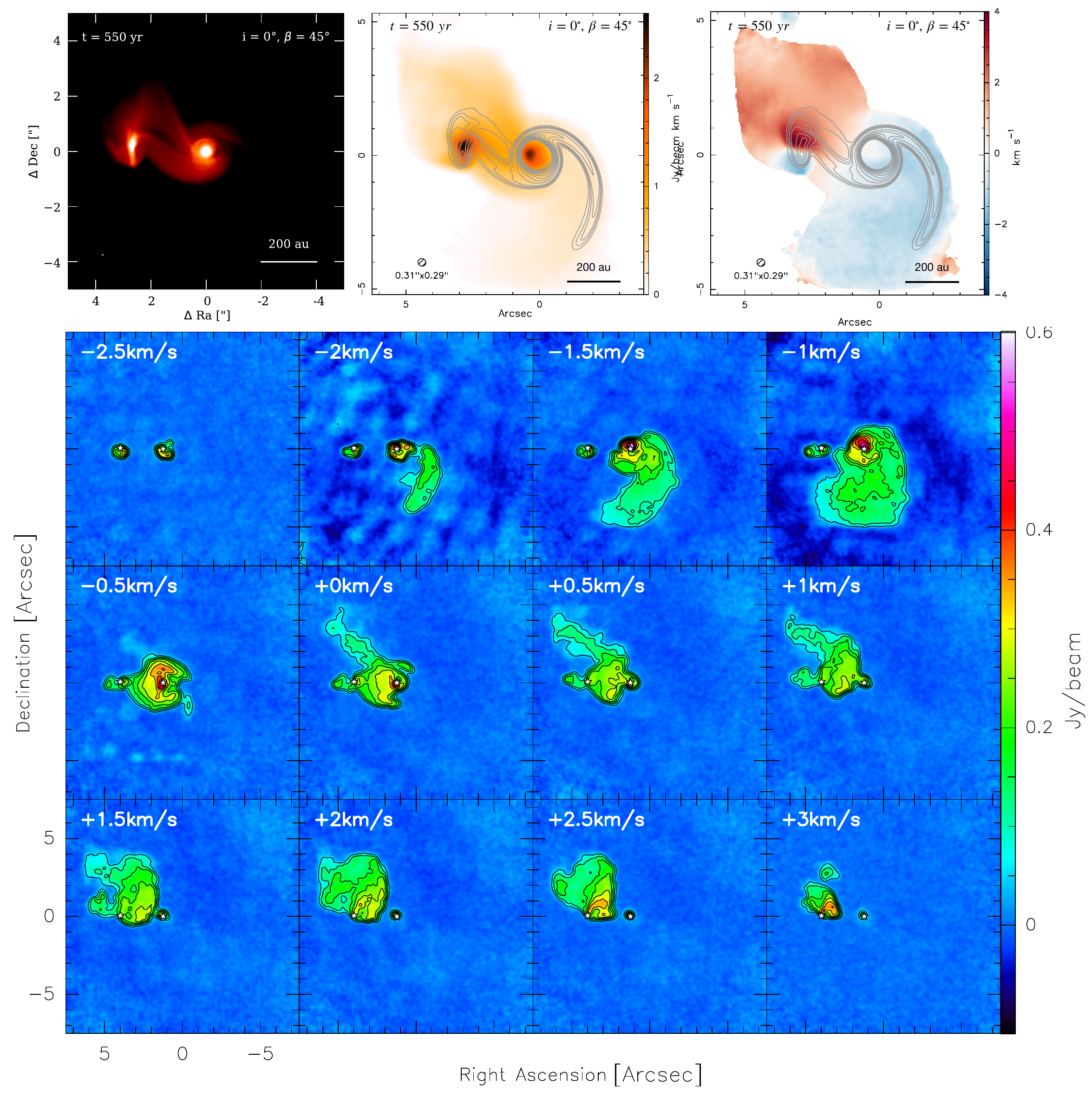}%
   \caption{\rr{Observations at different wavelengths for the inclined prograde $\beta=45\degr$ encounter, taken 550 yr after the passage at pericentre. Upper row: scattered light from Fig.~\ref{fig:scatteredlight-pro} (left);} moment 0 (middle), and moment 1 (right) of the $^{12}$CO(3-2). The contours in grey highlight the continuum emission at 850~$\mu m$ (same as the top-left panel in Fig.~\ref{fig:continuum}). Bottom: channel maps of the $^{12}$CO(3-2). The contour levels in each panel start at 5$\sigma$ with 5$\sigma$ steps, where $\sigma=10.5$\,mJy/Beam. \rr{The white stars indicate the location of the stars.} The non-coplanar southern spiral clearly appears in the negative channels between $-2$ and $-0.5$~\kms. The perturbed disc around the primary and the bridge of material in between the two stars is seen for channels around the systemic velocity ($0\pm 1.0$~\kms). This is in agreement with the scattered light emission (Figs.~\ref{fig:scatteredlight-pro}.}%
    \label{fig:CO-channels}%
\end{figure*}

Figure~\ref{fig:CO-channels} shows \rr{the synthetic observations at different wavelengths for the inclined prograde $\beta=45\degr$ encounter taken $550$~yr after the passage at pericentre. In particular, in the lower part we show} the $^{12}$CO(3-2) moment~0 (left) and moment~1 (right) maps. Moment~0 provides information about the distribution of gas around both stars. As already seen in scattered light images, the {\it bridge} of material connecting both stars is readily seen in CO moment~0. The western spiral is also detected out to separations of $5$ arcsec, \rr{which is roughly twice the pericentre distance}. Because of illumination effects and obscuration this prominent spiral feature is not detected in scattered light images. \rr{However, faint spirals could be detected in the CO \citep{Christiaens+2014}. Hence, CO emission lines can be used to reveal material spread around the stars during and after flybys --- not seen in scattered light.}

The moment~1 map in Fig.~\ref{fig:CO-channels} provides information about the velocity field around each star in the rest frame of the host star. Within the field of view, any non-coplanar disc with respect to the plane of the sky appears as an almost symmetric region with a given spread in velocities (typically of a few km/s). Hence, the features in the right panel of Fig.~\ref{fig:CO-channels} reveal evidence of both discs. The disc around the primary (although more massive and extended) has a relatively weak observational signature in moment-1 because it is almost coplanar to the plane of the sky. The disc of captured material around the perturber, being more inclined, has a larger kinematical signature, see middle column of Fig.~\ref{fig:scatteredlight-pro}. Prominent disc structures (such as spirals) translate into more asymmetric patterns in the moment 1 map. These features are however better seen by scanning through the individual channel maps.

Figure~\ref{fig:CO-channels} shows the $^{12}$CO J=3-2 channel maps at 0.5~km/s resolution (from +3 down to -2.5 km/s). The spiral in the South-West appears prominently across most of the negative channels from -1 up to -2.5 km/s. The width of the spiral decreases with increasing channel velocity (from systemic velocity). This is because the channels corresponding to faster velocities (e.g. -2 km/s) trace the lower disc surface while the channels for slower velocities (e.g. -1 km/s) trace the bulk of the disc instead.

The presence of the spiral across a broad range of velocities is a clear signature of non-coplanarity --- see discussion in Sect.~\ref{subsec:FUOri} for examples where such structures have been detected. Between -1 and +1 km/s we see the rotation pattern of the disc around the primary plus the asymmetries due to the presence of the spiral. Since this disc is inclined only by a few degrees the classical `butterfly pattern' \citep[e.g.][]{Louvet+2018} does not appear cleanly.

The disc around the perturber is inclined by approximately $\sim$80$\degree$ with respect to the plane of the sky. This explains why its rotation pattern is readily seen between +1 and to +3 km/s --- a coplanar disc with the plane of the sky does not produce a rotation pattern in the moment~1 map. We notice that half of the butterfly pattern appears in the positive channels (from +1.5 up to +3 km/s) where there is little or no overlap with the emission of the circumprimary disc. The interstellar bridge appears at velocities of about $1$~km/s. These kinematic signatures along with the dust continuum emission constrain the mutual inclination between the two discs. We discuss this further in Sect.~\ref{sec:flybyobs}.

\section{Discussion}
\label{sec:discussion}

\subsection{When should you suspect a flyby?}
\label{sec:suspect}

The four main observational signatures of a stellar flyby in a protoplanetary disc are:

(i) {\it Spirals:} for the prograde configurations two prominent spirals appear in the disc \citep{Clarke&Pringle1993, Pfalzner2003, Quillen+2005}. Because of projection effects and disc stripping one of these might appear as a bridge connecting both stars (see middle column of panel a in Fig.~\ref{fig:scatteredlight-pro}). For retrograde flybys, less prominent spirals form for misaligned orbits and almost no spirals at all for coplanar orbits (Fig.~2 in Paper~I). When spirals appear, these axisymmetric features efficiently trap dust. This trapping can be seen and quantified through multi-wavelength observations in scattered light (Fig.~\ref{fig:scatteredlight-pro}) and in the continuum (Fig.~\ref{fig:continuum}). Spirals are expected to disappear over time as the disc recircularizes (in a few thousand years or less). As seen in fig.~2 of Paper~I, the pitch angles of the two diametrically opposed spirals are not necessarily identical, and evolve over time from a few tens to several degrees. The bridge is out of the disc plane and fades more rapidly than the opposing spiral, which remains coplanar with the disc. The spirals that appear during the encounter last for a few thousand years (at most) for the parameters considered here. Hence, if spirals are observed in the continuum this means that the encounter must be ongoing or that the perturber is at a distance of a few thousands of au from the disc-hosting star. Given the probability of witnessing a flyby, this remains unlikely but possible nonetheless (see Sect.~\ref{sec:statflybys}).

(ii) {\it Disc truncation:} prograde flybys result in efficient disc truncation unless $R_{\rm peri} \gg R_{\rm out}$ or $q\ll1$ \citep{Clarke&Pringle1993, Ostriker1994, Breslau+2014, Winter+2018a}. Instead, retrograde encounters cause little or no disc truncation at all (fig.~11 in Paper~I), unless the flyby is penetrating enough ($R_{\rm peri} \leq R_{\rm out}$). We find that the dust distribution in the disc is more compact than the gaseous distribution due to radial drift\footnote{This is especially true for the grains marginally coupled to the gas, typically of sizes ranging from 0.1 mm up to 1 cm \citep{Laibe+2012}.} \citep{W77}. This difference in radial extent increases with time. Therefore, during a flyby, the gaseous disc should in principle show more structure than the dusty one. In regions of high stellar density, discs can also be rapidly truncated by external photoevaporation as shown by \citet{Winter+2018b}. This is further discussed in Sect.~\ref{sec:statflybys}.

(iii) {\it Disc warping:} a disc is referred to as warped when the angular momentum of the gas changes as a function of radius, described by the tilt and twist angles \citep{Pringle1996}. Prograde flybys are less efficient than retrograde ones regarding disc warping (sect.~3.8, figs.~13 and D6 in Paper~I). Remarkably, retrograde inclined orbits ($\beta=135\degr$) cause the strongest disc tilting \citep{XG2016} because the stripping is less severe and more material survives at larger radii. This effect is also apparent in Figs.~\ref{fig:scatteredlight-pro}, ~\ref{fig:scatteredlight-retro}, and \ref{fig:continuum} when comparing $\beta=45\degr$ with $\beta=135\degr$. Finally, the longevity of the warp generated by the flyby depends on the disc thickness and the outer radius of the disc \citep[e.g.][]{Nixon&Pringle2010}. Even after the warp dissipates the disc will maintain its misalignment to the central star and --- if observed --- this could be interpreted as a signpost of a previous encounter. In that case, high-precision radial measurement of the nearby stars (as the ones obtained with {\it Gaia}) could help to chase the hypothetical perturber. Larger perturber masses (i.e. $q>1$) produce more significant disc warping.

(iv) {\it Diffuse halo and captured material:} during a flyby, disc material can remain bound to the primary, be captured by the perturber or become unbound. For a given value of $R_{\rm peri}$, the process of disc stripping is more dramatic for prograde encounters and for high values of $q$. This phenomenon can be detected through molecular line emission, as shown in Fig.~\ref{fig:CO-channels} with the $^{12}$CO J=3-2 emission. For prograde (close enough) encounters, disc material is efficiently captured by the perturber. Alternatively, perturbers on retrograde orbits hardly steal material from the disc, unless $R_{\rm peri} \lesssim R_{\rm out}$. If both stars have discs previous to the encounter, then a complex exchange of material can happen between both discs. This dynamical effect is out of the scope of the present work, but has potentially been observed in AS~205 for instance (see Sect.~\ref{subsec:AS205}). 

\rr{(v) {\it Dimming and extinction events:} The material that is spread around the stars during the encounter can also cause extinction. The column density along the line of sight (or equivalently the bolometric stellar flux) over time is strongly dependent on flyby parameters, disc structure, and viewing angle. This renders the dimming signature of a flyby highly degenerate. In RW~Aur, such dimming events have been observed over the last years \citep{Gunther+2018}. For instance, in the model proposed by \cite{Dai+2015}, the extinction is due to the flyby-induced tidal arm (see their figure 12). Depending on the position of the observer, the diffuse halo around the stars, the tidal arm, or the bridge of material can cause similar dimming events. This signature, although not unique and ambiguous, provides a straightforward way to identify potential systems of interest within ground-based surveys of stars.}


\subsection{Prograde or retrograde flyby?}
\label{sec:provsretro}

In Table~\ref{tab:signatures} we summarise the flyby diagnostics presented in Sect.~\ref{sec:suspect}. In particular we separate each specific signature at various wavelength in prograde and retrograde cases (P and R, respectively). This provides a guide to interpret observations where a flyby is suspected. The presence of an interstellar bridge along with a diffuse halo is a signature of a prograde flyby; whereas a significantly misaligned disc is more indicative of a retrograde flyby. Multi-wavelength observations (e.g. scattered light, dust continuum, and emission lines) are crucial to disentangle between both orbital orientations.

\begin{table*}
\begin{center}
\caption{Observational signatures of flybys for different kind of observations: scattered light, dust continuum, and emission lines (e.g. CO). The symbols $\sim$ and $\times$ express a less robust diagnostic and the lack of information, respectively. \rr{These flyby signatures are ranked by relevance. Other mechanisms can produce similar signatures, except for bridges.}}
\label{tab:signatures}
\begin{tabular}{|c|c|c|c|}
\hline
Signature & Scattered Light & Dust Continuum & Emission Lines \\
\hline
    Bridges & Prograde & Prograde & Prograde \\
    Spirals & Prograde, $\sim$Retrograde & Prograde, $\sim$Retrograde & Prograde, $\sim$Retrograde \\
    Disc misalignment & Retrograde, $\sim$Prograde & Retrograde, $\sim$Prograde & Retrograde, Prograde \\
    Disc truncation & Prograde & $\sim$Prograde & $\sim$Prograde \\
    Accretion event & $\sim$Prograde & $\times$ & $\sim$Prograde \\
    Dust trapping & Prograde, $\sim$Retrograde & Prograde, $\sim$Retrograde & $\times$ \\
    Diffuse halo & Prograde, $\sim$Retrograde & $\sim$Prograde & Prograde, $\sim$Retrograde \\
\hline
\end{tabular}
\end{center}
\end{table*}


\subsection{What can we learn from the kinematics?}
\label{sec:warps}

\begin{figure}%
    \centering
   \includegraphics[width=0.5\textwidth]{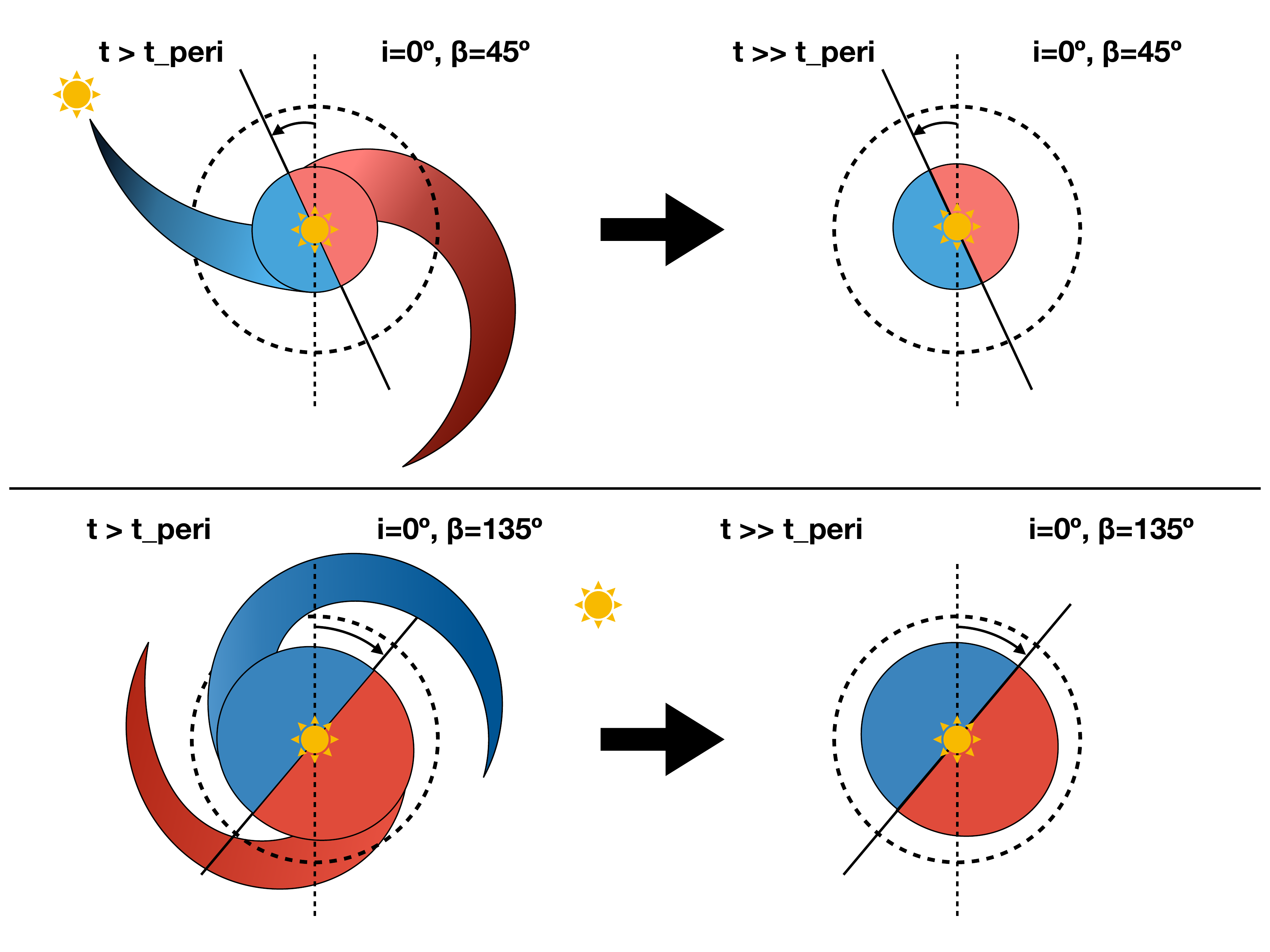}%
   \caption{\rr{Disc kinematics during an inclined prograde flyby ($\beta=45\degr$, top row) and an inclined retrograde flyby ($\beta=135\degr$). The disc is shown shortly after the passage at pericentre ($t>t_{\rm peri}$, left) and once the spirals and bridges have totally disappeared ($t \gg t_{\rm peri}$, right). The material in blue (red) is moving towards (away from) the observer. The colour gradients correspond to the velocity departure from the stellar systemic velocity: dark and bright correspond to fast and slow velocities. The velocity gradient increases with distance to the central star (assumed to be at systemic velocity). The dashed black line shows the radial extend of an unperturbed disc. Perturbers on prograde and retrograde orbits twist the disc in counter-clockwise and clockwise directions, respectively. Retrograde (prograde) perturbers are more efficient in tilting (truncating) the disc.}}
    \label{fig:kinematics}%
\end{figure}

In Sect.~\ref{sec:CO} we described the kinematical signatures of a stellar flyby as seen in the $^{12}$CO(3-2) emission line. In particular, the moment~1 map provides information about disc rotation and orientation. Assuming Keplerian rotation around each star, it is possible to obtain an estimate of the stellar masses even in the case of strongly embedded objects.

If the spectral resolution is high enough (ideally above 1~km/s), gas flowing out of the disk can be separated from the bulk of the disc. In particular, the prominent spirals generated by an inclined prograde perturber appear as arc-like features across several individual channels as shown in Fig.~\ref{fig:CO-channels}. In a configuration in which the disc is mildly inclined with respect to the plane of the sky, these two arc-like features would appear in the red-shifted and blue-shifted channels with respect to the v$_{\rm lsr}$ of the disc. More interestingly, if the blue-shifted (respectively red-shifted) channels were stacked together, the arc-like features would translate in a conical morphology. In V2775~Ori, \citet{Zurlo+2017} interpreted a double cone as an evidence for a bipolar outflow, when it could be generated by two spirals out of the disc mid-plane (similar to $\beta=45\degr$).

\rr{In Fig.~\ref{fig:kinematics} we show a sketch of disc kinematics during inclined flybys as a function of time (as seen in moment 1 CO maps for instance). The disc is initially face on and does not have any kinematical signature. Shortly after the passage at pericentre ($t>t_{\rm peri}$, left), the disc is warped and prominent spirals appear in the disc for $\beta=45\degr$ and $\beta=135\degr$. In the prograde case, the spirals are more radially extended and one of the spirals is seen as a bridge of gaseous material between the two stars. These spirals are not coplanar with the disc and hence have velocity departures of several km/s with respect to the systemic velocity. Spectral resolutions of the order of $1$ km/s or even higher are necessary to properly map the vertical layers of the disc. Perturbers on prograde and retrograde orbits cause twist in different directions. Moreover, retrograde flybys are more efficient in tilting the disc; whereas prograde ones cause more dramatic disc truncation. These kinematical signatures are key to reconstruct the geometry of the encounter.}

For a prograde flyby where disc stripping occurs and material is captured by the perturber, it is in principle possible to reconstruct the flyby geometry based on a single observation. Assuming there was no disc around the perturber prior to the encounter, the disc rotation pattern observed around the perturber constrains the orbital inclination. The moment~1 map is particularly useful in this regard to measure the relative orientation between discs. However, the problem often encountered when modelling interacting objects is that it is hard to infer the disc radial extent and disc alignment prior to the flyby. Several observations during the encounter would be ideal to restrain the (likely broad) range of possible flyby and disc configurations. However these might span over several decades or even centuries, which renders this task challenging.

Lastly, twisted isophotes \rr{in channel maps} and rotated structures in the velocity field as in \citet{Rosenfeld+2014} and \citet{Walsh+2017} could indicate the presence of a warp in the disc. Although this has been mainly applied to circumbinary discs as HD~142527, flybys should generate similar kinematical signatures in the first moment maps of $^{12}$CO (J=3-2, J=6-5) and HCO+ (J=4-3) for instance.


\subsection{When is a flyby statistically likely?}
\label{sec:statflybys}

Close encounters ($R_\mathrm{peri}\lesssim 100$~au) between individual stars are statistically rare in regions with stellar densities $\lesssim 10^4$~stars/pc$^{-3}$, typical of the vast majority of star forming regions in the solar neighbourhood \citep{Winter+2018b}. However, as discussed in Sect.~\ref{sec:flybyobs}, a number of such encounters have been inferred in local environments. 

The resolution to this apparent paradox is that approximately half of all stars form in multiple systems \citep{Raghavan+2010}. In some cases this can result in a stable binary which can influence disc evolution \citep{Papaloizou+1977,Daemgen+2013,Kurtovic+2018}. In other cases, the decay of higher order multiplicity \citep[or scattering of a third star due to the large cross-section of a binary -- e.g.][]{Hut+1983} can lead to the chaotic ejection of individual stars, during which particularly close encounters are possible. This is likely the case for HV~Tau~C and DO~Tau \citep{Winter+2018c}. In support of this hypothesis, \citet{Kraus+2008} find evidence of a structured spatial distribution of stars on length scales $\gtrsim 0.04$~pc in Taurus. This structure is not found on smaller scales, where it may have been erased by dynamical interactions between stars. Such a scenario is consistent with hydrodynamic simulations of star formation, which predict that encounters between stars in multiple systems are common in the early stages ($\lesssim 0.1$~Myr) of cluster evolution \citep{Bate2018}. 

Since multiplicity does not appear to be strongly dependent on environment \citep[see][for a review]{Duchene+2013}, it follows that the influence of dynamical encounters on disc evolution is similarly independent of stellar clustering. This explains the apparent high occurrence rate of star-disc encounters in low density star forming regions. Counter-intuitively, in massive and dense clusters external photoevaporation by FUV photons rapidly depletes the disc from the outer edge, such that the truncated disc may actually be less likely to be influenced by dynamical encounters \citep{Johnstone+1998,Adams+2004, Facchini+2016,Haworth+2018,Winter+2018b}. Hence, while encounters do not represent an environmental mechanism for disc dispersal, they play an important role in setting their initial conditions in all stellar birth environments. They also occur almost exclusively in the early stages of cluster evolution, such that the chance of observing individual cases is low. This further motivates the theoretical exploration of encounter signatures presented here so that the small number of observed cases are understood as such. 

\subsection{Observed flyby candidates}
\label{sec:flybyobs}

The signatures above-mentioned (see Table~\ref{tab:signatures}) are useful to interpret systems where an ongoing (or past) flyby is suspected. We note that the repeated interaction with a bound companion generates similar dynamical signatures but more compact spirals and discs. Below we discuss a few systems of interest.

\subsubsection{RW~Aur}
\label{subsec:RWAur}

RW Aur is a system composed of two stars: RW Aur~A and RW Aur~B with masses of 1.4~$M_\odot$ and 0.9~$M_\odot$ \citep{Ghez+1997}, respectively. The presence of a prominent tidal arm observed in CO in the disc around RW Aur~A \citep{Cabrit+2006} strongly suggests that RW Aur~B is perturbing the disc. \cite{Dai+2015} self-consistently modelled this system through hydrodynamical simulations considering a parabolic ($e=1$), inclined ($\beta \sim 20\degr$), and prograde encounter with $q\approx0.64$. Moreover, RW Aur~A is observed to have a high accretion rate ($\sim 10^{-7} \, M_{\odot}$/yr, \citealt{Hartigan+1995}), consistent with a prograde encounter.

More recently, \cite{Rodriguez+2018} reported new observations of RW~Aur in the continuum and in $^{12}$CO J=2-1, at higher resolution and for larger field of view. The dust discs around both stars appear symmetrical given the beam sizes and shapes. Also, based on the presence of additional tidal streams, the authors suggest that the RW~Aur system has undergone multiple flyby interactions. The radial extension of most prominent tidal arm is puzzling since several flybys would have heavily truncated the disc. Also, the likelihood of experiencing several stellar flybys during disc evolution is extremely low. \rr{In addition, several optical dimming events have been reported between 2011 and 2017, see \cite{Gunther+2018} for instance. These authors also report a sudden increase in Fe abundance during the event seen in X-ray emission. This feature is difficult to explain with a stream of gas passing by at a large distance. They suggest it is caused by the collision of (iron-rich) planetesimals close to the star. This would be a direct effect of the increase in eccentricity in the disc due to the perturber RW Aur B.}

\subsubsection{AS~205}
\label{subsec:AS205}

AS 205 is multiple stellar system where two components have been resolved at 168~au projected separation. AS 205~N is a pre-main sequence star with a mass 0.87 $M_\odot$, and AS 205~S is a spectroscopic binary with a total mass 1.28 $M_\odot$. The latest observations of this system reported by \cite{Kurtovic+2018} show two compact discs in the continuum, one around AS 205~N and one around AS 205~S. They also reported extended emission of gas subject to complex kinematics between the two systems using $^{12}$CO(2-1) emission. Remarkably, there is a bridge of gas between both sources and two spirals in the dust disc around AS~205~N. This feature strongly suggests that we are witnessing a prograde flyby with $q\approx1.5$, where the two discs are interacting. Finally, a spiral pattern appears in the channel maps of the CO emission at around $4$ km/s.

\subsubsection{HV~Tau \& DO~Tau}

HV Tau is a triple system with a wide binary, HV Tau~C, at $\sim 550$~au projected separation; and a tight binary with $10$~au separation. HV Tau~C hosts a protoplanetary disc and it is separated by $\sim$1.26$\times~10^4$~au from DO Tau, which also hosts a disc. There is a clear bridge between both source in the 160~$\mu$m emission. \citet{Winter+2018c} modelled the interaction between HV~Tau~C and DO~Tau as the decay of a quadruple system. In particular a penetrating disc-disc prograde encounter is required to unbind sufficient mass to produce the visible bridge. The mass ratio of the components is quite unconstrained, but an equal-mass encounter ($q\sim1$) is within errors. HV~Tau~C additionally exhibits a high accretion rate \citep{Woitas&Leinert1998}, also suggestive of a prograde flyby.

\subsubsection{FU~Ori, V2775~Ori, and Z~CMa}
\label{subsec:FUOri}

Accretion (or equivalently outburst) events and prograde encounters are deeply connected as originally proposed by \citet{Bonnell&Bastien1992}. This mechanism could explain the stellar brightness increase of 5 or 6 mag observed in FU Ori objects. FU~Ori itself is a multiple system where FU~Ori~N and FU~Ori~S have masses of 0.3~$M_{\odot}$ and 1.2~$M_{\odot}$ \citep{Beck&Aspin2012}, respectively. FU~Ori~N hosts a disc and exhibits a high accretion rate that reaches values as high as of $10^{-4}\,M_{\odot}$/yr. Both stellar components show compact discs in the continuum. There is also a prominent spiral out of the FU~Ori~N disc-plane (P\'erez et al., subm.), which coincides with the spiral seen in scattered light \citep{Takami+2018}. This evidence strongly supports the idea that FU Ori is indeed a system experiencing a dramatic prograde flyby ($q\approx$4).

Other recently imaged discs such as Z~CMa \citep{Takami+2018}, V2775 Ori \citep{Zurlo+2017}, and V1647 Ori \citep{Principe+2018} exhibit suspiciously similar disc structures, along with outburst events. In Z~CMa there is a prominent and open spiral arm in the disc, which could be explained by a prograde flyby in the past (Dong et al., in prep.). In V2775~Ori, in order to explain the peculiar CO emission lines, \citet{Zurlo+2017} proposed a ``double cone outflow''. A prograde stellar flyby instead provides a natural explanation for the arc-like features observed in the kinematics (see for instance Fig.~\ref{fig:CO-channels}). Therefore, we suggest that a fraction of the FU Ori-like objects might be experiencing a flyby. This can be confirmed with higher resolution observations.


\subsection{Caveats}
\label{sec:caveats}

The synthetic observations shown in this work correspond to the specific case of a parabolic ($e=1$) non-penetrating flyby ($R_{\rm p} > R_{\rm out}$) between two solar-mass stars ($q=1$). Moreover, we only considered two inclinations of $\beta=45\degr$ (prograde) and $\beta=135\degr$ (retrograde). The reason why we focus on parabolic encounters is two-fold: first, as we discuss in Section~\ref{sec:statflybys}, dynamical encounters between unrelated stars are rare such that the majority of star--disc encounters are expected to occur during the early stages of cluster evolution between (proto)stars in multiple stellar systems. In such interactions, encounters are by definition gravitationally focused,\footnote{i.e. the stars have low relative velocities at infinity.} such that $e\approx 1$. Second, star--disc encounters for which $e\approx 1$ induce the greatest angular momentum transfer and therefore generate the most prominent structures in the disc \citep{Vincke&Pfalzner2016,Winter+2018b}. More specifically, unbound perturbers on hyperbolic trajectories ($e>1$) translate into encounters at higher velocities where the mechanism of eccentricity excitation within the disc is less significant \citep{Winter+2018b}.

When calculating the kinematics, we choose a face-on view to emphasise the features generated by the non-coplanar structures identified in our simulations. A consequence of this choice is a weak signal from the primary disc, seen in Figure~\ref{fig:CO-channels}. However, we note that even small deviations from this particular orientation will result in a measurable signal from the disc --- e.g., as in TW Hya which is misaligned to the viewer by only $4\degr$ \citep{Huang+2018,Andrews+2018,Flaherty+2018}.

Lastly, we note that we set the distance of our system at 100~pc from Earth. This is a somewhat optimistic value since the most studied star forming regions are at distances ranging from 140~pc up to 400~pc. Therefore, discs in these regions would appear smaller and have lower resolution in the observations. However, even at larger distances, prominent signatures as shown in Sect.~\ref{sec:synthobs} should be readily observed.

\section{Conclusions}
\label{sec:conclusions}

Flybys produce remarkable and distinctive observational features at different wavelengths. By combining multi-wavelength observations it is possible to reconstruct an observed flyby (perturber's orbit, disc geometry, mass ratio, and pericentre distance). Scattered light and emission lines are particularly efficient at probing the gas distribution around each of the stars and their potential misalignment. We regard these to be the most powerful diagnostics for detecting flybys. Dust continuum adds information about the flyby impact parameters. 

The main observational signatures of flybys are summarised in Table~\ref{tab:signatures} and are the following:
\begin{enumerate}
  \item Spirals and bridges: These are identified clearly in scattered light observations (Figs.~\ref{fig:scatteredlight-pro} and \ref{fig:scatteredlight-retro}) and are more prominent for prograde encounters. Such asymmetries efficiently trap dust (Fig.~\ref{fig:continuum}). Additionally, misaligned encounters also leave non-coplanar kinematic signatures (e.g. CO channels, Fig.~\ref{fig:CO-channels}).
  \item Warps and disc misalignment: Particularly for retrograde flybys, disc warping is observed in moment 1 maps (Fig.~\ref{fig:CO-channels}). Once the warp dissipates, the disc is expected to remain misaligned with respect to its host star.
  \item Disc truncation: A more compact dust disc than gas is recovered in the observations (Fig.~\ref{fig:continuum}). Prograde encounters more severely truncate the disc.
\end{enumerate}

Our catalogue of synthetic observations of two representative flybys provides a way to interpret recent observations of multiple objects where a flyby is suspected (see Sect.~\ref{sec:flybyobs}: \rr{RW~Aur, AS~205, HV~Tau \& DO~Tau, FU~Ori, V2775~Ori, and Z~CMa}). Finally, the lack of bound companions in some non-axisymmetric systems --- despite an active search --- could be well explained by a past stellar flyby. Future observations will help to better estimate the occurrence of such encounters and understand the subsequent process of planet formation in these discs.


\section*{Acknowledgements}

We thank the anonymous referee for useful suggestions. NC acknowledges financial support provided by FONDECYT grant 3170680. NC and JC acknowledge support from CONICYT project Basal AFB-170002. FL acknowledges the support of the FONDECYT program n$^\circ$ 3170360. This project has received funding from the European Union's Horizon 2020 research and innovation programme under the Marie Sk\l{}odowska-Curie grant agreement No 823823 (DUSTBUSTERS). CP and DJP acknowledge funding from the Australian Research Council via FT170100040, FT130100034 and DP180104235.
FMe acknowledges funding from ANR of France under contract number ANR-16-CE31-0013.
AJW, RN, GD, and RA acknowledge financial support from the European Research Council (ERC) under the European Union's Horizon 2020 research and innovation programme (grant agreement No 681601). JC and MM acknowledges support from Iniciativa Cient\'ifica Milenio via the N\'ucleo Milenio de Formaci\'on Planetaria. MM acknowledges financial support from the Chinese Academy of Sciences (CAS) through a CAS-CONICYT Postdoctoral Fellowship administered by the CAS South America Center for Astronomy (CASSACA) in Santiago, Chile. LC acknowledges financial support provided by FONDECYT grant 1171246. The Geryon2 cluster housed at the Centro de Astro-Ingenier\'ia UC was used for the calculations performed in this paper. The BASAL PFB-06 CATA, Anillo ACT-86, FONDEQUIP AIC-57, and QUIMAL 130008 provided funding for several improvements to the Geryon/Geryon2 cluster.

\newpage
\bibliographystyle{mnras}
\bibliography{flybybiblio}

\appendix

\end{document}